# THE DEVELOPMENT OF ELECTRONIC PAYMENT SYSTEM FOR UNIVERSITIES IN INDONESIA: ON RESOLVING KEY SUCCESS FACTORS

Veronica S. Moertini[1], Asdi A. Athuri[2,4], Hery M. Kemit[3], Nico Saputro[1]

[1]Informatics Dept., [2]Accounting Dept., [3]IT Bureau, [4]Finance Bureau
Parahyangan Catholic University
Bandung – Indonesia
`moertini, asdi, kemit, nico@home.unpar.ac.id`

*ABSTRACT*

*It is known that IT projects are high-risk. To achieve project success, the strategies to avoid and reduce risks must be designed meticulously and implemented accordingly. This paper presents methods for avoiding and reducing risks throughout the development of an information system, specifically electronic payment system to handle tuition in the universities in Indonesia. The university policies, regulations and system models are design in such a way to resolve the project key success factors. By implementing the proposed methods, the system has been successfully developed and currently operated. The research is conducted in Parahyangan Catholic University, Bandung, Indonesia.*

## 1. INTRODUCTION

The trend shows that more and more Indonesians do financial transactions electronically via banks ATMs, internet banking and SMS-banking. It was reported that in an Indonesia major bank, the number of ATM transactions is 3.5 to 4 million (of total 5 million transactions) in a single day [1]. In another bank, the ATM transactions are increased by 32% annually [2]. Still, the banks in Indonesia seem to be confident that the ATM transactions would keep on increasing in the next several years as they continue to add more ATMs in the entire areas of Indonesia [3] [4]. As reported in [5], currently, web technology has also been used widely. In line with this trend, the internet transactions in Indonesia are also increased fantastically (up to 87% in a single quarter) so is the SMS and mobile banking transactions [6, 7, 8, 9, 10]. With this trend, it could be said that conducting transactions electronically has been becoming necessity and way of life of Indonesians.

By considering the trend, it is necessary that higher education institutions (HEIs) in Indonesia, where most depend on the students' tuition to operate, provide electronic payment services for the students. Since students in an HEI might origin from all over area of Indonesia, and their tuitions might be paid directly by their parents, it would be best if the payment system could be accessed through the banks wide ATM network, internet, and cellular network established in

International Journal of Computer Science & Information Technology (IJCSIT), Vol 3, No 2, April 201117

Indonesia. What a pity, though, the result of our survey on the HEI websites reveals that among more than 3000 HEIs in Indonesia [11], only less than ten have been operating "online system payment" where most of their online systems accept payment transactions conducted in limited or a few banks network only. Certainly, this could be considered as a drawback in terms of catching up with IT use and fulfilling the Indonesians needs of electronic transactions. By conducting limited survey on a few HEIs, we also found the fact that the financial HEI managements nowadays still encounter problems related to balancing the total of student bills with the actual payment made. They are also having difficulties in generating the up to date yet accurate financial reports. These, of course, cause inefficiencies in the HEIs.

Developing Electronic Payment System (EPS) to handle tuition bills and payment is a complex task, since students' and parents' behavior as well as their needs, HEI management requirement, integration with the academic information system, IT infrastructure, banks products offered and banks collaboration must be addressed. The task is also challenging as the literatures acknowledge that IT projects are high-risk. (It was reported that 34% of IT projects were late or over budget, 31% were abandoned, scaled back or modified, only 24% were completed on time and on budget [12]. [13] also reported that only 28% of software development projects were successful.)

Contributions: Despite the urgent needs of EPS for HEIs and the challenge in developing one, we could find research results of EPS concerning with e-commerce or government institutions only (see Literature Study section). Therefore, this paper is intended to contribute in presenting the development method and model of EPS for universities (a HEIs type with the largest number in Indonesia), which specifically ensure the project success. Our approach is: We define the key success factors in the early stage of requirement gathering, then, in the following stages, the university policies and system models are designed in such a way to ensure that key success factors are achieved. As the banks maintain confidentiality of their system for the sake of security, detailed discussion of the system development could only be presented for the system component owned by the university. The research is conducted in a well known university in Indonesia, Parahyangan Catholic University (Unpar), which currently employs about ten thousands undergraduate and graduate students.

The paper is presented in the following order: Introduction, literature study of related research result, key success factors for the electronic payment system, the proposed methods of resolving each of the key success factors, result and evaluation, and conclusion.

## 2. LITERATURE STUDY

### 2.1. Electronic Payment System (EPS) Criteria

An electronic payment is defined as a payment services that utilize ICT, including cryptography and telecommunications networks [14]. EPS is classified into cash-like systems (e-cash), check-like systems (credit card and credit-debit based systems), and hybrid systems (stored-value card based systems) [15]. [16] states that in order to be successful in implementing EPS, users' awareness must be increased, users are encouraged to use it, and be assured that the system is secure and comprehensive. Infrastructure to accommodate high quality telecommunication facilities must also be provided. [17] presents findings from a survey to identify customers' attitude towards a new payment service, which are: usage (score: 0.990), ease-of-use (0.922), usefulness (0.898) and behavioral intention to use (0.889). [18] and [14] present the significant criteria in system evaluation, which are security, cost, convenience and universality (includes payment type and interoperability across systems). [14] suggests that to increase customer interest, they should be able to choose the payment instrument with the lowest cost. Customers must also be able to keep track on the balance. Based on these findings, it could be summarized that in designing EPS, the most important issues that must be addressed are efficiency, security, convenience, cost, flexibility or universality, privacy, reliability, customer interest, and infrastructures.

## 2.2. Related Research Result

To of our knowledge, there is no similar research result that covers all the aspects that we would discuss. Most of the research results in the area of EPS are concerned with e-commerce. Nevertheless, the following are some that are related to our proposed model.

[19] presents a per-fee-link framework that is independent of the payment protocol, which could be used in e-commerce, and is specially designed to support micro-payments. Two main actors participate in this framework, which are a client using his browser and the vendor's web server. Additionally, a payment service provider (an entity that accepts payments on behalf of the vendor) could participate depending on the payment protocol used to make the payment. After selecting goods (such as MP3 player, PDF reader, etc.) online, the client could complete the payment (for one item only or several items - depending on the business model designed) via browser using the suitable protocol.

[20] discusses a multi-user electronic bill presentment and payment model that enhance the current billing systems as well as overcome the problems of periodic generation of billing reports and mail volumes. The biller sends the detailed bills to the consolidation website and the summary to the customers. The registered customers (of the consolidation website) could check and pay the bills at any time. Here, the consolidation website could have many registered-biller clients.

[21] presents a pilot project to determine the feasibility of adopting an application service provider solution to support procurement by multiple federal agencies using a variety of different legacy transaction systems. The Internet Payment Platform (IPP) by the eMoney group of the US Treasury's Financial Management Service involved three federal agencies and subsets of their suppliers. As the system deals with a variety of different legacy transaction systems, the proposed IPP system integrates the agencies MIS and supplier systems by interfacing. The agencies utilize the integrated system to perform goods procurement while the suppliers could further process the transaction.

## 2.3. IS Project Quality Standards and Risk Management

In managing information system (IS) projects, the quality assurance process comprises of all the activities and actions required to ensure that the project meets the quality standards outlined during the quality planning phase [22]. One of the quality standards is the project quality metrics, which must be met at the project completion. In this context, managing risks could be viewed as a series of actions to ensure that the metrics are met. Although the role of developer team in ensuring project' success is very significant, most literatures (such as [12], [23] and [13]) present research findings in the area of risk management from the point of view of the clients or management. Fortunately, [24] stressed the significant of managing risks from the from the developers' perspective and proposed the methods accordingly. However, [24] does not specifically discuss how to ensure that the project standards are met.

The risks factors of the projects could be related to people, process, technical issues and so on, which could emerge throughout the project lifecycle, namely, initiation, development and implementation (see [12, 23, 25]). The strategies for dealing with risks could be classified into risks avoidance, reduction and mitigation [13, 22].

## 3. THE KEY SUCCESS FACTORS (KSFs)

One of the steps for managing project quality is defining the project quality metrics [22] that should be achieved at the project completion. Here, we define the metrics in the form of KSFs, which are defined to address the problems faced by the university managements, problems faced by and culture of the students and parents, as well as to meet the EPS criteria as described in Section 2 (more detailed discussion in designing the KSFs has been presented in our previous research result in [26]). As the whole system will integrate the university system (**US**) and bank partner system (**BS**), here, we identify which system should resolve each of the KSFs (see Table 1).

Table 1. KSFs, targets and the system resolving the KSF.

| No | KSF Definition | Target | Resolved by |
|---|---|---|---|
| KSF-1 | The cost of infrastructure and information system development is affordable. | Less than USD 15,000. | BS-US |
| KSF-2 | System maintenance could be handled by one staff/operator. | Less than 10 hour/week of staff working hour. | BS-US |
| KSF-3 | Additional charge for online payment | Zero. | BS |
| KSF-4 | System availability & easy access. | Anywhere, anytime (multi-bank payment). | BS |
| KSF-5 | The users use the system. | 100% of students or parents. | BS |
| KSF-6 | Number of maximum payment transactions that could be handled in a day. | All users make the payment transaction anytime at the given period of time. No fail transaction. | BS-US |
| KSF-7 | Bill statement accuracy. | 100% at any time. | US |
| KSF-8 | Transaction accuracy (exact payment). | 100% at any time. | BS-US |
| KSF-9 | Balancing between bills and payments | Guaranteed 100%. | BS-US |
| KSF-10 | Payment transaction data transfer. | Real time (time lag is less than 30 sec.). | BS-US |
| KSF-11 | Partial payment. | Two payment transactions / semester. | BS-US |
| KSF-12 | Bills are collected and transactions are conducted by students or parents who have the right to do so. | No breach, no complaint from users. | BS |
| KSF-13 | Provide or guarantee identification, confidentiality, authentication, data integrity, non-repudiation, customer solvency, and durability. | All of the criteria achieved. | BS |
| KSF-14 | Provide real time bill and transaction reports. | Criteria achieved. | US |

## 4. METHODS FOR RESOLVING THE KSFs

In our previous research result [26], we present the proper stages for electronic payment system for universities development, which are: Cultural issue and problems identification, system solution definition, bank partner selection, requirement elicitation, system model and design, implementation, test and deployment. In each stage, the strategies for dealing with risks that we adopt could be classified into risks avoidance and reduction, which are appropriately addressing the KSFs given in Table 1.

In this section, we present the risks avoidance and reduction implementation in the analysis and design stage (for database designed only), which we view as the most important ones in ensuring project success. The material that would be discussed are the methods for resolving the KSFs in the bank partner and payment product selection, the university policies and regulations (in the requirement elicitation stage), designing each of the system model (architecture, data-exchange and interaction between systems, statechart, as well as data flow diagram) and database design.

### 4.1. Bank Partner and Payment Product Selection Criteria

In selecting the bank partner and its payment services, the KSFs defined in Table 1 that should be resolved by the bank system are "translated" into the more meaningful criteria for the selection. As there are lots of banks operated in Indonesia where each offers a few payment products, we designed two stages of selection. The first is the banks candidate selection and the second is payment product selection, each stage use a set of different criteria given in Table 2. This way, the detailed survey for evaluating payment products is performed for a few banks passing the first selection only.

Table 2. The criteria to resolve KSFs for selecting the bank partner and payment product.

| Selecting | KSF Resolved | Criteria |
|---|---|---|
| Bank partner | KSF-5, -6, -8, -13 | Have millions of customers and a fine reputation for its security and reliability in handling millions of customer accounts and transactions. |
| | KSF-4 | Have many branches throughout the country and establish a network that integrates the branches. |
| | KSF-4 | Have a secure established communication network with other major banks that facilitates money transfer (across banks) through ATMs, e-banking and SMS banking. |
| | KSF-4 KSF-5 | Offer payment services that facilitate payment anywhere anytime, as well as exact payment with minimal charges. |
| Payment Services | KSF-9, -11, -12, -13 | Offer a payment product receiving only the exact amount as stated on the bill for each student. |
| | KSF-1 KSF-2 KSF-10 | Offer real time data transfer of payment transaction through free leased line. Offer no charge for the EPS development and establishing the communication to the university system. The maintenance of the bank system is conducted by the bank. |
| | KSF-3 | Offer no extra charge for students as well as the university on every tuition payment transaction made. |

In the selection process, we found that the banks in Indonesia offer 3 payment products for companies or institutions, which are Virtual Account (**VA**) with Open Payment (**VA-OP**), Host to Host with Exact Payment (**H2H-EP**), and Virtual Account with Exact Payment (**VA-EP**). The winner is **VA-EP** offered by Bank X having the properties as follows: (1) It provides a unique account for each student, which is used for tuition payment destination. As a regular transaction account, it accepts payment transactions processed by Bank X as well as other bank through various services and devices (bank's teller, ATMs, clearing, SMS-banking, e-banking, etc.) (2) It only accepts payment with the amount as stated in the tuition bill (which is sent by the university system to VA-EP system).

### 4.2. University Policies and Regulations

As stated in [25] and [23], one of the major risk factors that fails IS projects is the users unfamiliarity or unfulfilled user needs. To manage this, the university policies must be designed such that EPS would be effectively used by users. Usually, in developing information systems for enterprises, the organization policies and regulations are defined at the beginning of the requirement elicitation stage, as they will be referred throughout this stage. However, we found that in developing the EPS which integrates two large systems (the external bank system and internal university system), the university policies and detailed regulations could only be defined once the bank partner and its unique payment services have been chosen. As student tuition bill generation and payment handling are the work of academic and financial management, the policies and regulations should come from these divisions.

The following are the university financial policies and regulations resolving several of the KSFs:

(1) **KSF - 7, 11**. Student bills are computed based on the most up to date of course registration data and scholarships. For regular semester, student bills are generated twice (before each semester begins and in the middle of the semester), while for short semester, the bills are generated once (after courses registration period).
(2) **KSF-8, -9**. For regular semester: (a) Bill-1 contains the registration fee and total tariff of 10 course credits (applied to undergraduate students) or 5 credits (applied to graduate students). Bill-1 is due on a certain date before course registration at the beginning of each semester. (b) Bill-2 contains the total bill for the semester (computed by summing up registration, credits of courses, labs and assistantship fee) deducted by the amount of Bill-1 and scholarship given (by the university or other institutions), if any. Bill-2 is due on a certain date before the mid-semester exam. For short semester: Bill-3 contains the total bill for the semester and due on a certain date before the mid-semester exam.
(3) **KSF-9**. If any bill is not paid before the due date, fine is computed for each bill, namely Fine-1, Fine-2 and Fine-3. If students could not meet the payment of the on going semester, the total bill and fine would be summed up and stated as Due-bill.
(4) **KSF-14**. The bill and fine items should be available online and could be accessed anytime and anywhere as the reference of payment.
(5) **KSF-3**. Tuition payment transaction is free of charge.

To put the financial regulations into effect, the academic regulations must also be designed in line with them. Hence, **KSF-9** is also supported by the following regulations:
For regular semesters: (1) At the beginning of each semester, students who have not completed Due-bill and Bill-1 are not allowed to register courses. Those who want to delay the payment would have to follow certain procedures to obtain approval of payment delay. 2) In the middle of each semester, students who have not completed Bill-1 and Bill-2 are not allowed to attend mid-term exams. Those who want to delay the payment would have to follow certain procedures to obtain approval of payment delay. (3) At the end of each semester, students who have not completed the total bill would not receive and not be able to view their grades online.
For short semester: (1) In the middle of short semester, students who have not completed Bill-3 are not allowed to attend course exams. (3) At the beginning of the following semester, students who have not completed the total bill would not receive and not be able to view their grades online.

### 4.3. System Architecture

The university policies and regulations could be imposed effectively if the bank partner's VA-EP system (to be brief, we would denote it as VAS), university payment system and the Academic MIS are integrated so that data integrity and consistency across the systems could be maintained in a real time basis. The architecture of the integrated system is given in Figure 1. The tasks of each system are:

(1) Academic MIS (**AMIS**): Sending the most recent data of the active students, course registration and scholarship, also receiving bill and tuition reports from UPS.

(2) University Payment System (**UPS**): Computing bills and communicating to VAS, processing payment and reversal request and generating reports.

(3) Virtual Account System (**VAS**): Handling various complex tasks, such as managing student account, payment transactions, communicating with other banks systems, clearing, reconciliations, etc.

In terms of resolving KSFs, the advantages of the proposed integrated system are: (1) At anytime, student bills could be computed precisely based on the current student data registration and scholarship received, then sent to VA System in real time basis (resolving **KSF-7 to -11**). (2) Academic sanctions could be imposed accordingly based on real time student bill payment status (**KSF-9**). (3) Students could pay tuition through many devices as well as view reports at anytime from anywhere (**KSF-4, -12**). With this architecture, the cost of infrastructure and information system development would also be affordable (**KSF-1**) as the communication link between the systems would use the existing infrastructure. The UPS maintenance could be handled by one operator (**KSF-2**) as most of the time the system would process transactions automatically and VAS would be managed by the bank partner. Hence, this architecture addresses all of the KSFs that could be resolved by IT system.

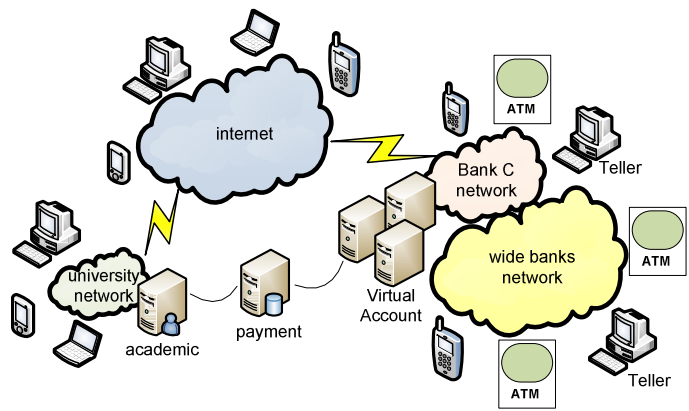

Figure 1. Payment system architecture that incorporates three large systems.

### 4.4. Data Exchange between the Systems

The system context diagram or Data Flow Diagram (DFD) level 0 is a necessary tool for developing a baseline interaction between systems and actors, actors and system or systems and systems [27]. The objective of the diagram is to focus the attention on external entities and events that should be considered in developing a complete set of system requirements and constraints. Here, as shown in Figure 2, UPS would exchange data with 2 other external systems and 2 groups of user. (The users and data flow of the external system – AMIS and VA System, are given with the intention to show the users' role in providing and receiving data used by and produced by UPS.) The significance of the interactions in addressing KSFs is given in Table 3.

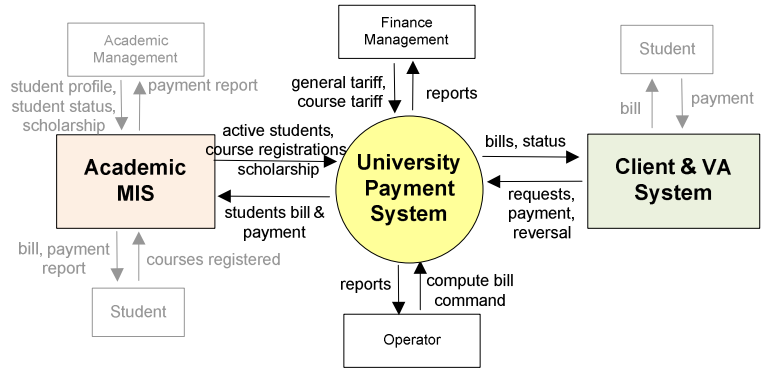

Figure 2. The diagram context of the whole system.

Table 3. The interactions between systems to resolve KSFs.

| KSF | Data Flow | Discussion |
|---|---|---|
| KSF-7 | From AMIS to UPS, data sent: active students, course registration, scholarship. From Finance Management to UPS, data sent: general tariff, course tariff. From Operator to UPS, data sent: compute bill command. | To guarantee the accuracy of the bills, data needed to compute precise bills for each active student is sent (from AMIS to PS) whenever data stored in AMIS are changed. Also, by supplying the current tariffs, Finance Management helps to guarantee that the bills generated are accurate. |
| KSF-8, -9, -10, -11, -12 | From VAS to UPS, data sent: requests (bill/reversal), payment/reversal. From UPS to VAS, data sent: bills, status | By VAS requesting the bill first before the payment transaction could be made (where the amount of payment must be equal to one of the bills), it guarantees the transaction accuracy, balancing bills-payment made and partial payment. This also supports the access of bills by the authorized users only. |
| KSF-14 | From UPS to AMIS, Finance Management, Operator, data sent: student bill, payment, reports. | This guarantees that the system provide real time reports. |

### 4.5. Interaction between VA System and University PS

The behavior of the system could be modeled using UML diagram of Common Behavior, Collaborations, Use Cases, State Machines, and Activity Graphs. Collaborations diagram specifies the concepts needed to express how different elements of a model interact with each other from a structural point of view [28]. Sequence diagram is one of the collaboration diagrams that could be chosen in modeling the interactions. A sequence diagram shows an interaction arranged in time sequence. In particular, it shows the instances participating in the interaction by their "lifelines" and the stimuli they exchange arranged in time sequence.

Modeling the interaction between UPS and VAS is very significant as it presents the sequences of the message exchange. Only after this is done, other models of UPS could be designed accordingly. (However, this model must be designed by both the developer team and bank partner as it involves the operations of the bank partner system.) In Figure 3, Client is the bank partner VA System's client such as teller system, ATM, and other systems that handle the payment transaction. The KSFs addressed by this model is given in Table 4.

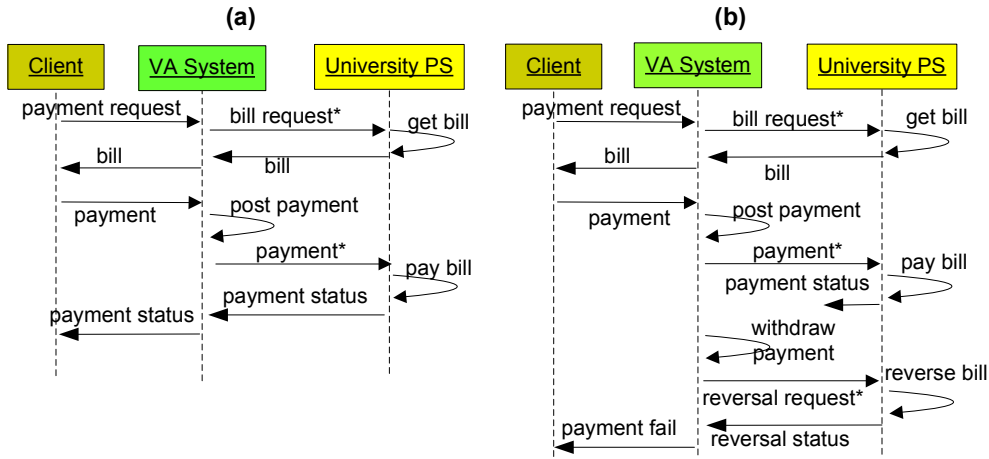

Note for Fig. 3: *: the message is sent up to 3 times to ensure it is gone through. Client: the system where transactions are made by students or parents (teller, ATM, SMS-banking, e-banking, etc.).

Figure 3. The sequence of message exchange between VAS and UPS, and the services executed: (a) for normal transactions (b) for transactions with reversal.

Table 4. The sequence of message exchange between VAS and UPS to resolve KSFs.

| KSF Resolved | Sequence of Messages | Discussion |
|---|---|---|
| KSF-7, -8, -9 | From Client to VAS, message: payment request, then from VAS to UPS, message: bill request*, from UPS to VAS, message: bill. From: Client to: VAS, message: payment, then from: VAS to: UPS, message: payment*, then from UPS to VAS, message: payment status. | The exchange of the messages following these sequences ensures that students will receive the correct bill to be paid. Then, by defining the sequence that payment could only be made after the bill is received, the system guarantees that the payment made is equal to the bill. (The bill request message could be sent up to 3 times to ensure that the request goes through.) |
| KSF-8, -13 | From: VAS to: UPS, message: reversal request then from UPS to VAS, message: reversal status | If the acknowledgment from UPS is not gone through (due to communication problems), VAS would ask for reversal for the payment that is just made. This guarantees the transaction accuracy and data consistency between VAS and UPS. |

## 4.6. UPS State Transition Diagram

A state transition (statechart) diagram can be used to describe the behavior of a model element such as an object or an interaction [28] or dynamic model of a system [29]. Specifically, it describes possible sequences of states and actions through which the element can proceed during its lifetime as a result of reacting to discrete events (e.g., signals, operation invocations). One of the behaviors of the system that could be described using statechart is the system operations. Typically, real time system behavior needs to be modeled using statechart.

While the sequence diagram models the interaction between UPS and VAS, the statechart is used to model the behavior solely in the UPS. The sequences of the UPS system operations should be arranged carefully in order to address several of the KFSs. The UPS statechart is shown in Figure 4, while the significance of sequence in addressing KSFs is given in Table 5.

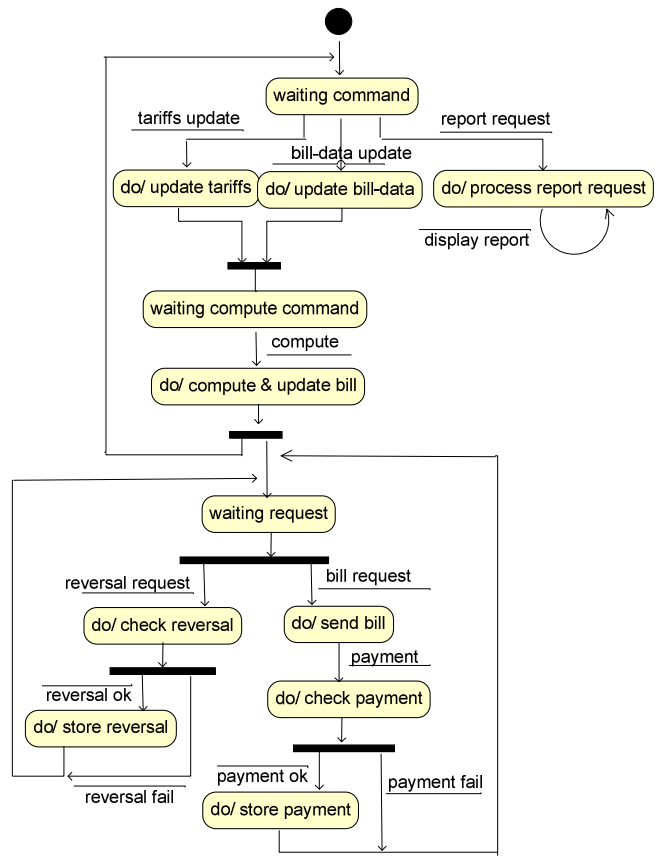

Figure 4. Statechart diagram of UPS operations.

Table 5. The sequence of states to address KSFs.

| KSF | Discussion |
|---|---|
| KSF-6 | UPS is ready at 24/7 to respond to bill request (from VAS) and then process the payment made. Since the complete transaction takes at most 3 seconds to process, the number of transactions that could be handled in one day is more than 28,000 transactions, which is more that the number of students. Therefore, this guarantees that all transactions made are handled. |

| KSF | Discussion |
|---|---|
| KSF-7, -8, -9, -10, -11, -13 | Bill request and payment could only be made after bill computation based on the most up to date data is done. Next, when a payment is received, UPS checks the amount. If the amount is correct, the payment is stored. However if reversal is requested from VAS, the system will process it and cancel the payment made. Hence, these sequences contribute in guaranteeing that bills are correct, any payment is matched to a bill, so that the sum of total bills is equal to the total of payment made. |
| KSF-14 | UPS is ready 24 hour in a day to receive and process report request to guarantee real time reports production. |

## 4.7. UPS Data Flow Diagram Model

Data Flow Diagram (DFD) is a modeling tool that could be used to model the information and functional domain at the same time [27]. DFD serves two purposes, which are to provide an indication of how data are transformed as they move through the system and to depict the functions that transform the data flow. There are several advantages of DFD, such as it clarifies the system processes, shows the sequence of the processes that have to be done in a particular order, identifies the name of data stores and files, etc. In terms of addressing KSFs, we find that this is a suitable model and the result is given in Figure 5 (the data dictionary of each data and storage seen in the DFD is given in Appendix A). The discussion of how the processes addressed KSFs is depicted in Table 6.

Table 6. The UPS processes to resolve KSFs.

| KSF | Resolved by Process | Discussion |
|---|---|---|
| KSF-6 | pay bill | pay bill is ready at any time to take the payment sent by VAS to guarantee that all payment transactions made are handled. |
| KSF-7 | update tariff update bill-data compute student bill get bill | Having the responsibility of taking and storing the data needed for bill computation, update tariff and update bill-data provide the most recent data to avoid false bills. Then, compute student bill guarantees the accuracy of the bills as they computed based on the most recent data. get bill, which stands by at anytime to process bill request coming from VAS, sends the most recent bill. By this mechanism, the bill statement accuracy is guaranteed. |
| KSF-8, -9 | pay bill | On processing each incoming payment transaction sent by VAS, pay bill checks whether it equals to one of the bills sent by get bill, and if the payment amount does not match any one, it will be refused. Therefore, it guarantees the balance between the bill and payment made. |
| KSF-10 | pay bill | pay bill, which stands by at anytime to handle payment transaction coming from VAS, promptly processes the payment to support speedy data transfer. |
| KSF-11 | pay bill | As pay bill accepts payment having the amount equals to one of the bills, which could be a partial bill, it guarantees that partial payment could be made |
| KSF-12, -14 | generate reports | generate reports produce the reports based on the most up to date data, display it to the authorized users and send it to AMIS in real time basis. Therefore, it guarantees the real time reports accessed by authorized users. |

| KSF | Resolved by Process | Discussion |
|---|---|---|
| KSF-13 | get bill<br>pay bill<br>reverse bill | Mainly, KSF-13 is resolved by the bank VAS. However, as get bill and pay bill conduct the tasks as described in KSF-7 to -10, while reverse bill processes the bill reversal requested by VAS (if the transaction is not completed due to communication problems), the three processes help to guarantee that KSF-13 is met. |

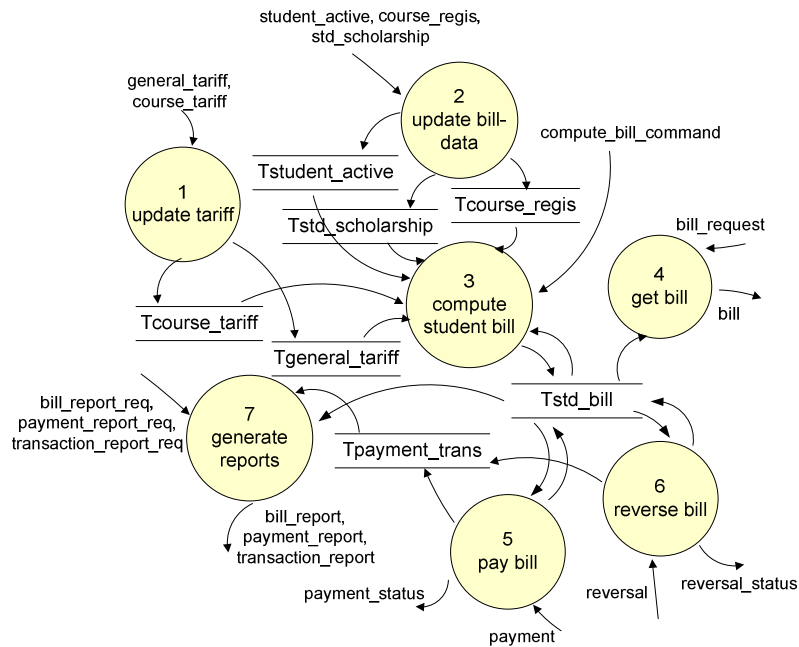

Figure 5. The data flow diagram of the UPS.

## 4.8 UPS Database Design

The methodologies in developing databases are conceptual, logical and physical designs [30]. Basically, the conceptual and logical purpose is to construct the models of information that guarantees the completeness, correctness, integrity and consistency of the data stored in the database. On the other hand, the physical aim is to design the implementation of the models that guarantee the security, scalability, and efficiency of the database.

In this context, the criteria applied to conceptual and logical design are translated into KSF-7 to KSF-12 and KSF-14 (see Table 7). The scalability (KSF-6) and efficiency such as speedy real-time transactions (KSF-10) and fast report generation (KSF-14) are realized by designing tables with no join operation, creating no index on the tables that are accessed in payment and reversal transactions (Tpayment_trans) but indexing on the fields that are frequently used as search keywords (year, semester, student_ID, tariff_ID, etc - see Appendix A) in the six other tables. The security of the database are realized by managing user privileges, roles, resources and auditing database use as discussed in [31, 32].

Table 7. The database tables to resolve KSFs.

| KSF | Resolved by Database Tables | Discussion |
|---|---|---|
| KSF-7 | Tstd_scholarship, Tgeneral_tariff, Tcourse_tariff, Tstudent_active, Tcourse_regis, | Used to store and maintain the most up to date data necessary for the bills production. |
| KSF-8, -9, 11 | Tstd_bill | Used to store the most up to date bills that should be matched by the payment. Each bill is stored as a table record and could be paid separately to guarantee partial payment. |
| KSF-10, -12, -14 | Tpayment_trans, Tstd_bill | Used to store the payment with speedy access and reports productions. |

## 5. RESULT AND TARGET EVALUATION

The proposed EPS has been developed, deployed and operated in Parahyangan Catholic University. It has been used to handle tuition payment for two regular semester (the 2$^{nd}$ semester of 2009/2010 and 1$^{st}$ of 2010/2011) and one short semester (July-August 2010). During the 3 semesters, 11.153 students and parents have utilized EPS successfully. The result in achieving the target defined for each KSF is given in Table 8.

Table 8. The target evaluation of EPS.

| No | Target | Result |
|---|---|---|
| KSF-1 | Less than USD 15,000. | Target is achieved. By utilizing the available network infrastructure, the total cost is approximately USD 13,000 (mainly used to pay the developer team salary). |
| KSF-2 | Less than 10 hour/week of staff working hour. | Target is achieved. UPS is maintained by one staff who spends approx 8 working hours/week. |
| KSF-3 | Zero. | Target is achieved. No transaction charge. |
| KSF-4 | Anywhere, anytime (multi-bank payment). | Target is achieved. Transactions could be conducted on ATMs, internet and smart phones. |
| KSF-5 | 100% of students or parents. | Target is achieved. All students use EPS for tuition payment. Total transaction made during 3 semesters: 31,938. |
| KSF-6 | All users make the payment transaction anytime at the given period of time. No fail transaction. | Target is approximately 99.5% achieved. The communication network and server could handle all transaction made as long as the electricity power is available (during the observation period, a few times the power was down (*)). |
| KSF-7 | 100% at any time. | Target is approximately 99.5% achieved. The problems: for special cases, where students are given academic or financial dispensations, the users (staff) sometimes are late in entering the data into the system, which caused the incorrect data needed to generate the bill. |
| KSF-8 | 100% at any time. | Target is approximately 99.99% achieved. Of all 31,938 transactions made, 38 are reversed. |
| KSF-9 | Guaranteed 100%. | Target is approximately 99.99% achieved. There are 6 |

| No | Target | Result |
|---|---|---|
| | | transactions that are recorded in the bank account but are not recorded in the UPS. |
| KSF-10 | Real time (time lag is less than 30 sec.). | No measured could be presented here. We encountered difficulties in measuring lag time between the transactions submitted (via ATMs, e-banking, tellers, etc.) to the UPS. However, all transactions made in VAS are sent to UPS in real time basis (data freshness is guaranteed), the communication between UPS and VAS through leased line is through 100 Mbps channel. |
| KSF-11 | Two payment transactions / semester. | Target is achieved. |
| KSF-12 | No breach, no complaint from users. | Target is achieved. |
| KSF-13 | All of the criteria achieved. | Target is achieved. No breach. |
| KSF-14 | | Target is achieved. |

(*) Note: during the EPS observation period, Indonesia was having electricity-power shortage that the electricity for companies and institutions sometimes were shutdown.

It could be seen in Table 8 that all of the targets are achieved or approximately achieved. We suspect that the problem in achieving the target of KSF-9 (achieved 99.99% only) is caused by the inter-bank payment transactions by clearing. The central-bank clearing system protocols might not comply with the protocols used by VAS causing the acceptance of the transactions with amount not equal to the bills by the bank partner (and stored in the student virtual account) but refused by the UPS.

## 5. CONCLUSION

The risks avoidance and reduction strategies that could be implemented to ensure IS project success is: Defining project KSFs at the beginning, then throughout the development stages each is resolved by designing the appropriate policies, regulations, system model and database tables design. Specifically for EPS development to handle tuition payment in the universities, which integrates the university system and bank system, the KSFs are addressed by both systems. Selecting the bank partner and its payment product, therefore, is a very important stage as several KSFs must be addressed by the two.

The detailed methods that we propose are specifically suitable to develop electronic payment system to handle tuition in the universities, with the case study of a university in Indonesia. They are also based on the classical paradigm of software engineering. In order to generalize the methods, larger scope of research is needed. The methods should also be enhanced using the newer paradigm of software engineering, such as object-oriented, service-oriented, etc., which are also suitable for IS development in organizations. It is also important that new or enhanced methods to address the main issues in electronic payment systems (efficiency, security, convenience, etc.), such as presented in [33], be implemented. Detailed discussion of the experiment results need to be presented as well to justify the proposed methods.

# APPENDIX A
The data dictionary of the UPS Data Flow Diagram.

The dictionary of the data that flow:
std_scholarship = year + semester + student_ID + name + scholarship_code + amount
student_active = year + semester + student_ID + pay_credits + bill1_credits
pay_credits = YES | NO
general_tariff = year + semester + tariff_ID + tariff_description + amount
course_tariff = year + semester + course_code + code_lab + amount_lab + code_studio + amount_studio + code_assist + amount_assist + code_tutor + amount_tutor
course_regis = year + semester + student_ID + name + course_code + credits + status_lab + status_studio + status_asist + status_tutor + trans_datetime
std_bill = year + semester + student_ID + paycode + amount + generate_datetime + paid_status + datetime_paid
bill_request = student_ID + transaction_no + bank_code + trans_datetime + del_channel + institution_code
bill = 00 + 360 + student_ID + {paycode + amount} (paycode and amount are repeated based on the content of Tstd_bill where paid_status = 'NO')
payment = transaction_type + student_ID + paycode + amount + CCY_code + bank_code + transaction_no + trans_datetime + del_channel + institution_code
payment status = SUCCESS | WRONG_AMOUNT | BILL_IS_ZERO | WRONG_ACCOUNT
reversal = transaction_type + student_id + paycode + amount + CCY_code + bank_code + transaction_no + trans_datetime + del_channel + institution_code
reversal status = SUCCESS | FAIL
bill_report = year + semester + student_ID + name + paycode + amount + generate_datetime + paid_status + datetime_paid
transaction_report = transaction_type + student_ID + paycode + amount + CCY_code + bank_code + transaction_no + trans_datetime + del_channel + institution_code
payment_report = student_ID + name + paycode + amount + trans_datetime
student = ALL | student_ID
acad_time = year + semester
compute_bill_command = {student_ID + paycode + acad_time}

The dictionary of the storages:
Tstd_scholarship = std_scholarship, Tgeneral_tariff = general_tariff, Tstudent_active = student_active, Tcourse_tariff = course_tariff, Tcourse_regis = course_regis, Tstd_bill = std_bill,
Tpayment_trans = payment, Tstudent_active = student_active

Where:
paycode = DUE-BILL | BILL-1 | FINE-1 | BILL-2 | FINE-2 | BILL-SS | FINE-SS | BILL-NS | TOTAL-BILL (where SS = short semester, NS = next semester)
tariff_ID = REGISTRATION_S1 | REGISTRATION_S2 | REGISTRATION_S3 | DEVELOPMENT_S1 | DEVELOPMENT_S2 | DEVELOPMENT_S3 | CREDIT_S1 | CREDIT_S2 | CREDIT_S3 (where S1 = undergraduate, S2 = master, S3 = doctorate degree)
CCY_code = currency (i.e. "IDR")

## Authors

Veronica S. Moertini is a lecturer in the Informatics Department, Parahyangan Catholic University (Unpar), Bandung, Indonesia. She completed her doctorate degree from Bandung Institute of Technology in August 2007. Her research interests include data mining, database and enterprise information systems. Moertini has published several papers in the conferences and journals in these research areas.

Asdi Athuri Aulia is a lecturer in the Accounting Department and the Head of Finance Bureau, Parahyangan Catholic University (Unpar), Bandung, Indonesia. His research interests include accounting, information systems, and so on.

Hery Maradona Kemit is currently a programmer in the Information Technology Bureau, Parahyangan Catholic University (Unpar), Bandung, Indonesia. He masters web-based, Visual Basic and database programming.

Nico Saputro is a lecturer in the Informatics Department, Parahyangan Catholic University (Unpar), Bandung, Indonesia. He is currently a doctorate student in the Department of Computer Science, Southern Illinois University, USA. His research interests include information systems, genetic algorithms, wireless mesh network, wireless sensor network, security and privacy, and Quality of Service routing.